\documentclass[sigconf,nonacm]{acmart}
\AtBeginDocument{%
  }

\newcommand{\squote}[1]{`#1'}
\newcommand{\myquote}[1]{\emph{``#1''}}
\newcommand{\code}[1]{\texttt{#1}}

\usepackage{multirow}
\usepackage{makecell}
\usepackage{enumitem}
\usepackage{placeins}
\usepackage{tabularx}
\usepackage{diagbox}

\begin{document}

\title{Augmenting Scholarly Reading with Cross-Media Annotations}


\author{Qi Xu}
\orcid{0009-0006-2125-3308}
\affiliation{%
 \institution{Web \& Information Systems Engineering Lab}
 \department{Vrije Universiteit Brussel, Pleinlaan 2}
 \city{Brussels}
 \postcode{1050}
 \country{Belgium}
}
\email{qi.xu@vub.be}

\author{Beat Signer}
\orcid{0000-0001-9916-0837}
\affiliation{%
 \institution{Web \& Information Systems Engineering Lab}
 \department{Vrije Universiteit Brussel, Pleinlaan 2}
 \city{Brussels}
 \postcode{1050}
 \country{Belgium}
}
\email{bsigner@vub.be}

\begin{abstract}
Scholarly reading often involves engaging with various supplementary materials beyond PDFs to support understanding. In practice, scholars frequently incorporate such external materials into their reading workflow through annotation. However, most existing PDF~annotation tools support only a limited range of media types for embedding annotations in PDF documents. This paper investigates cross-media annotation as a design space for augmenting academic reading. We present a design exploration of a cross-media annotation tool that allows scholars to easily link PDF content with other documents and materials such as audio, video or web pages. The proposed design has the potential to enrich reading practices and enable scholars to guide and support other researchers' reading experiences.
\end{abstract}



 \maketitle

\section{Background}
Annotations play a central role in supporting the reading of academic literature and have been shown to enhance scholarly reading in multiple ways. They facilitate navigation by directing readers' attention and reshaping the reading order. By treating annotated passages as salient and unannotated content as less critical, annotations guide readers toward focused reading of key content while enabling efficient skimming of less critical parts. This can improve both initial reading and subsequent rereading of scientific papers. Scim~\cite{fok_scim_2023} exemplifies this function by highlighting salient content to support skimming and help readers prioritise the most important information. Furthermore, annotations aid interpretation by providing explanatory content that enriches or clarifies the original text. Such explanations may include cross-lingual translations or plain-language summaries to support comprehension. Recent solutions like Paper Plain~\cite{august_span_2023} demonstrate this approach by providing in-situ, section-level summaries generated by large language models to support understanding. This interpretive role can be further extended toward semantic, machine-readable representations of scholarly content. Finally, annotations might provide support by linking to additional materials, including related articles and other media types such as videos. For instance, Papeos~\cite{kim_papeos_2023} enables scholars to link academic papers directly to their corresponding conference videos, enriching the reading experience with complementary materials.

However, annotations that provide support through additional materials remain insufficiently supported in current PDF-based scholarly reading tools. When engaging with a PDF article, scholars commonly rely on two types of external resources: related PDF documents (cross-document) and other media, such as web pages, images and videos (cross-media). Most mainstream academic PDF reading tools, including Google Scholar PDF Reader and the PDF~viewers embedded in citation management tools such as Zotero and Mendeley, support neither cross-document nor cross-media annotations. In practice, scholars are limited to embedding URLs within textual comments, which provides only a rudimentary form of linking. Some specialised tools, such as LiquidText~\cite{tashman_liquidtext_2011}, support cross-document annotations by linking to other PDF documents, but still lack support for cross-media annotations. More recent academic reading systems have begun to explore cross-media annotation as a promising research direction. Papeos~\cite{kim_papeos_2023}, for instance, showcases the potential of cross-media annotation by linking scholarly articles to conference videos, demonstrating how heterogeneous media can enrich comprehension and context.

Nevertheless, existing approaches typically focus on a single, predefined media pairing---such as the fixed document-to-video links provided by Papeos---rather than offering a general, extensible framework for cross‑media annotation across diverse resource types. This leaves a significant gap in current scholarly reading environments, where flexible support for integrating heterogeneous external materials remains largely absent.

As noted by Unsworth~\cite{unsworth2000scholarly}, scholars across disciplines increasingly engage with diverse forms of materials and frequently need to analyse and compare multiple objects of study simultaneously, whether these are texts, images, videos or other types of human-generated content. We argue that scholarly reading would significantly benefit from a generalised cross-media annotation approach that supports linking PDF documents not only to other PDFs but also to a wide range of scholarly and contextual resources, including audio, video and web pages. This motivation underlies our proposed design for a scholarly cross‑media annotation tool that enables researchers to create, access and integrate cross‑media annotations seamlessly within their academic reading experience.

\begin{figure*}[htb]
  \centering
  \includegraphics[width=\textwidth]{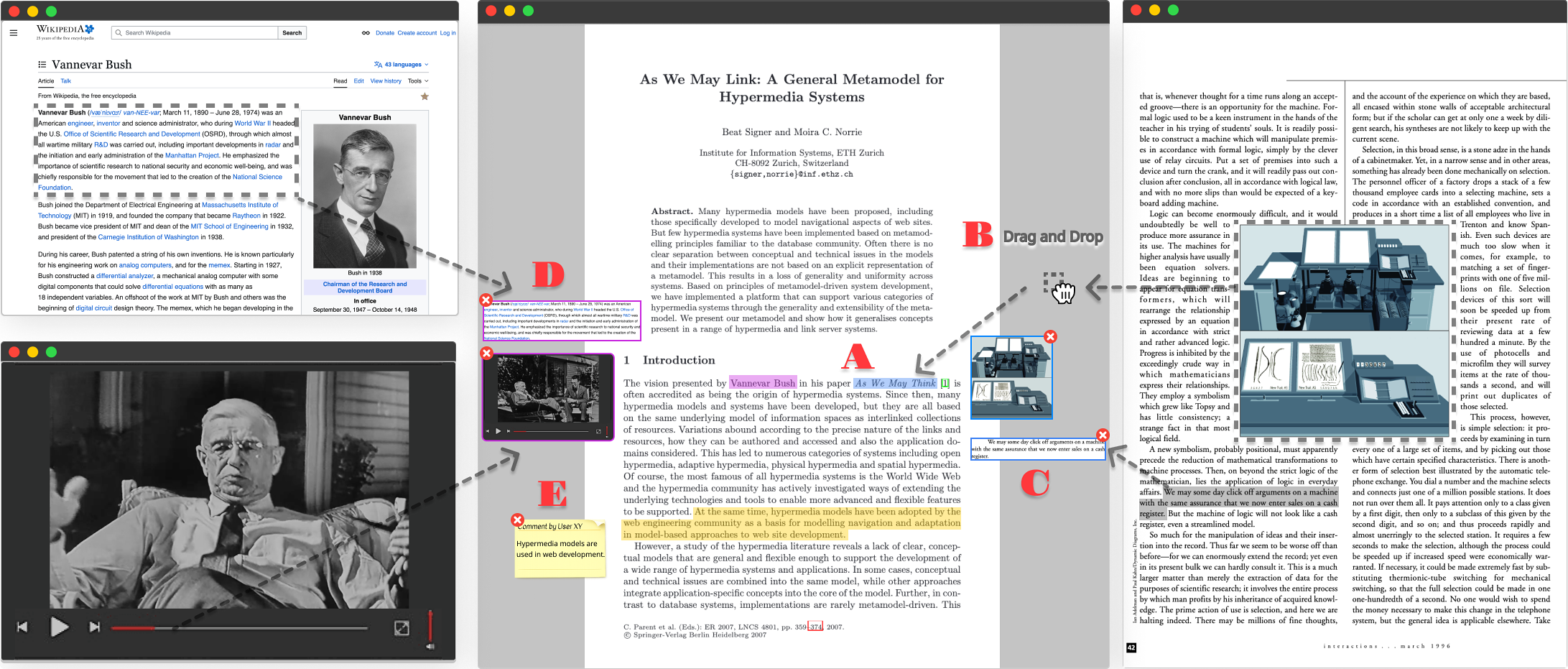}
  \caption{The central PDF viewer serves as the primary space for reading and annotation, while external materials consulted during reading are displayed on both sides (web pages and videos on the left, other PDF documents on the right). Scholars can highlight content in the main PDF that requires additional support~(A). By dragging a block or segment from an external resource onto the corresponding highlight, a pop-up widget representing the linked material is created in the PDF view~(B). A~single highlight may be linked to multiple external resources, forming one-to-many relationships~(C). The interface supports both cross-document and cross-media annotations, including connections to web pages and video segments~(D), as well as basic user-generated comment annotations~(E).}
  \label{fig:annotating}
\end{figure*}

\section{Cross-Media Annotation Tool Design}
We propose a novel cross-media annotation tool designed to support scholars in creating and viewing cross‑media annotations through simple and low-effort interactions, as illustrated in Figure~\ref{fig:annotating}. The design is guided by two core principles. First, while reading a PDF~document, scholars should be able to seamlessly link external resources as annotations to specific parts of the document. Second, they should be able to easily access external resources referenced in PDF annotations during the reading process. To illustrate how these principles shape our design, we present a representative usage scenario that demonstrates how scholars create and browse cross‑media annotations using the tool.

Let us consider a scholar reading a paper entitled \squote{As We May Link: A~General Metamodel for Hypermedia Systems}. While reading, the scholar encounters unfamiliar concepts and decides to consult additional external materials to deepen their understanding. In particular, the scholar has not previously read Vannevar Bush's seminal article \squote{As We May Think}, which is cited in the text. As depicted in \emph{step~A} of Figure~\ref{fig:annotating}, the scholar highlights the phrase \myquote{As~We May Think} in the current PDF document. The highlight colour, which appears to be blue in this case, is automatically assigned by the annotation tool as explained in more detail later. The scholar then finds a PDF~version of \squote{As We May Think} and identifies useful content within that document, as shown on the right-hand side of Figure~\ref{fig:annotating}, including both text and images. Using our proposed cross-media annotation tool, the scholar links these external materials back to the original document. As illustrated in \emph{step~B}, the scholar selects a region of the external PDF containing an image and drops it onto the blue-highlighted text in the main document, thereby creating an annotation. As a result, the PDF~viewer automatically generates a pop-up widget that displays the linked material in the margin adjacent to the highlighted text, using a proximity-based placement strategy. The widget's frame is further colour-coded in blue, matching the corresponding highlight and making the association visually explicit. As illustrated in \emph{step~C}, the scholar may also select text spans from external documents and link them to the main PDF~document using the same drag-and-drop interaction. This example demonstrates that a single highlighted item can be linked with multiple target annotations, enabling one-to-many relationships to be explored during reading. If the scholar later decides to remove an annotation, it can be deleted by clicking the control~(x) located in the top corner of the corresponding pop-up widget.

\begin{figure*}[htb]
    \centering
    \includegraphics[width=0.88\textwidth]{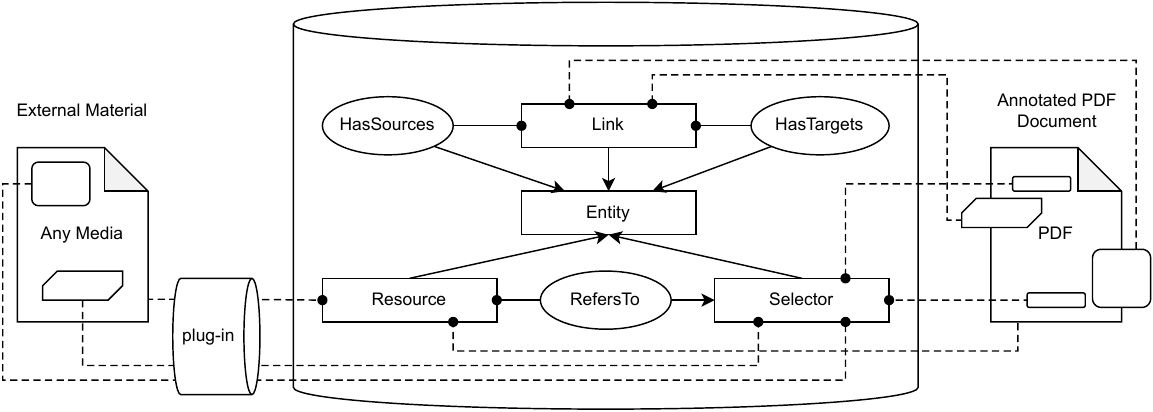}
    \caption{Cross-media annotation tool architecture}
    \label{fig:architecture}
\end{figure*}

Up to this point, the scenario has illustrated cross-document annotations. However, our tool further generalises this interaction to support cross-media annotations. As the scholar continues reading, they become interested in learning more about Vannevar Bush. To obtain additional contextual information, the scholar consults online resources and discovers both a relevant web page and a video interview with Vannevar Bush. As shown in \emph{step~D} of Figure~\ref{fig:annotating}, the scholar can select a paragraph from the web page and a segment of the video, linking each to the current document using the same interaction pattern previously introduced for PDF‑based materials. The linked content is then presented through pop‑up widgets, enabling the scholar to view both the web page excerpt and the video segment directly within the reading environment. Finally, as illustrated in \emph{step~E}, the cross-media annotation tool also supports basic comment annotations. Comments are handled consistently with other external resources and are stored as resources that are associated (linked) with the corresponding highlighted content.

To facilitate effective browsing of cross-media annotations, our design adopts a combined spatial and visual encoding strategy. The proposed design effectively leverages the grey margin space that naturally appears when the PDF viewer is zoomed out, allowing scholars to conveniently browse linked cross‑media content without leaving the document view. To further help users easily identify which highlighted passage each pop‑up annotation corresponds to, the interface redundantly employs two visual channels: colour hue and spatial position. Pop-up widgets are placed adjacent to their associated highlights using a proximity-based strategy, while linked content is rendered with an outline in a colour that matches the colour of the corresponding highlight. As noted earlier, highlight colours are automatically assigned by the tool. The system uses a palette of up to twelve distinguishable hues, assigning them sequentially from top to bottom in the document view and reusing them cyclically once all twelve distinct hues have been allocated.

\section{Architecture}
The architectural foundation of our proposed cross-media annotation tool builds on the general cross-media annotation framework introduced by Decurtins~et~al.~\cite{decurtins_putting_2003}. This framework conceptualises annotations as a distinct class of links within an open hypermedia system and emphasises media independence, extensibility and fine-grained addressability across heterogeneous resources. In particular, it uses the resource-selector-link~(RSL) hypermedia metamodel~\cite{signer2007we,signer2019} as its core data abstraction. In our work, we build on this general cross-media annotation framework to design a specialised cross-media annotation architecture tailored to PDF-based scholarly reading. Note that we also plan to support different semantic classifications of specific annotations, for instance, indicating whether an annotation is a comment, an explanation or an example, or to distinguish between formal and informal annotations~\cite{decurtins_putting_2003}.

In the architectural overview provided in Figure~\ref{fig:architecture}, the frontend components are presented on the left- and right-hand side. Since our design requires cross-media annotations to be rendered in situ within the PDF document during reading, a PDF rendering library with strong interactive support is essential. We plan to adopt the Semantic Reader~\cite{lo_semantic_2023} project's PDF component library\footnote{https://github.com/allenai/pdf-component-library} as the foundation for PDF rendering. A central strength of this library is its well-designed overlay layer, which manages most interactive elements, including page highlights and citation popovers. We plan to extend this overlay layer to support the creation and visualisation of cross-media annotations created by our tool.

The rendering and interaction of external materials are delegated to specialised desktop applications. For instance, web pages are displayed in a web browser and videos are handled by a video player. We obtain the information required to create cross-media annotations, such as selected fragments and metadata, by interfacing with these applications via lightweight plug-ins, following techniques previously proposed for cross-media document linking and navigation~\cite{tayeh_cross-media_2018,tayeh2014}.

The central part of Figure~\ref{fig:architecture} illustrates how scholars’ annotations are structured and stored in the system’s database. The backend is implemented as a REST server that exposes the functionality of the RSL~hypermedia metamodel. Within this model, all annotation-related elements are represented through three core entity types: \code{Resource}, \code{Selector} and \code{Link}. In our context, a \code{Resource} represents a scholarly artefact, such as a PDF document or an external media item. When a scholar opens a new PDF document in the frontend, the system automatically creates a corresponding \code{Resource} entity in the backend. A \code{Selector} identifies a specific fragment of a \code{Resource}, such as a text span, a region selected within a PDF~page or a segment of a video clip. When a scholar highlights content in the PDF viewer, a \code{Selector} entity is created and connected to the associated resource via a \code{RefersTo} relation. Similarly, for external materials, the entire item is stored as a \code{Resource}, while metadata about selected parts is managed by the corresponding \code{Selector} entities. Note that Figure~\ref{fig:architecture} presents a simplified view of these concepts, while a more comprehensive description of the RSL~hypermedia metamodel can be found in~\cite{signer2007we}.

Cross-media annotations between selected content are represented as \code{Link} entities. When scholars create connections between items via drag-and-drop interactions, the frontend issues requests to the backend to instantiate the corresponding \code{Link}. Each link maintains explicit \code{HasSources} and \code{HasTargets} relations to the associated entities, which may be either resources or selectors in our case. It is important to note that no content is duplicated. Instead, selected fragments of external resources are transcluded via the flexible, bi‑directional links offered by the RSL metamodel. When a scholar later reopens the same PDF document, the frontend automatically retrieves all related resources, selectors and links for the given PDF~resource. This enables the reconstruction and in‑situ rendering of previously created cross‑media annotations.

\balance

\section{Discussion and Conclusion}
The presented design offers three key benefits for scholarly reading. First, it supports scholars during the initial reading of an article by enabling them to quickly collect, organise and store external materials related to the current PDF document. By allowing external resources to be directly linked to highlighted content, the system helps scholars consolidate relevant information in context. This contextual integration accelerates sensemaking and fosters deeper comprehension during early-stage exploration.

Second, it supports the creator of annotations during rereading by providing them with a richer reading experience. By placing cross-media annotations directly alongside the corresponding highlighted content in the PDF, the system enables scholars to revisit a document together with the related materials they previously curated. Keeping linked targets visible in situ through margin-based placement improves recall and encourages deeper re-engagement when rereading scholarly articles. 

Third, the design supports shared and collaborative annotations. More experienced scholars can share their annotated documents with other scholars, such as early-career researchers or PhD candidates, allowing the linked external materials, in the form of annotations, to guide readers in understanding key concepts, methods and debates within a field’s literature. In summary, our proposed design supports scholarly reading across multiple stages by enabling the efficient collection of supporting materials during initial reading, recalling prior insights and deepening understanding during rereading and helping other scholars more rapidly familiarise themselves with the relevant literature of a given field through shared, context‑rich annotations.

While the current design centres on the manual and explicit creation of cross-media annotations, in which scholars actively search for and select external materials to link to a PDF, future developments may incorporate AI-assisted support, combining human-authored annotations with automatically generated recommendations. With the increasing adoption of large language models in scholarly workflows, researchers are relying more on AI-based solutions to retrieve and summarise relevant information rather than conducting all searches manually. Building on this trend, we envision integrating a recommendation mechanism into the cross-media annotation tool to automatically suggest and generate cross-media annotations that align with the content of a PDF~document. These AI-assisted annotations would not replace human judgment but rather serve as an optional, augmentative layer that enhances comprehension and supports exploratory reading. 

In this paper, we investigated cross‑media annotation as a design space for augmenting the scholarly reading process. Moving beyond traditional document‑bound annotation practices, we conceptualised annotations as cross‑media links that connect heterogeneous resources across documents and media types. Our proposed cross‑media annotation tool enables scholars to flexibly embed materials from multiple external sources directly into a PDF document, thereby enriching the document with contextual information. At~the same time, these cross‑media annotations are reintegrated into the reading environment in a lightweight, unobtrusive manner through margin‑based placement and selective visual encodings, ensuring that the enriched content remains accessible without disrupting the reading flow.

We regard the presented work as an initial step toward rethinking annotations as a cross-media scholarly practice. The proposed solution for augmenting scholarly reading with cross‑media annotations also forms part of our broader research effort on cross-media information management for next-generation scholarly workflows~\cite{xu2026}. Future work includes developing a fully functional implementation of the proposed solution and conducting empirical studies through both controlled experiments and in-the‑wild deployments to evaluate how scholars integrate cross‑media annotations into their reading, sensemaking and collaboration practices. Beyond validating usability and user experience, such studies may surface new patterns of cross‑media engagement that can further inform tool design. Additional avenues include exploring AI-assisted cross-media annotation recommendation mechanisms and investigating collaborative scenarios in research teams.

\begin{acks}
The work of Qi Xu is supported by the China Scholarship Council~(CSC) through a doctoral scholarship (Grant No.~202408330076).
\end{acks}

\bibliographystyle{ACM-Reference-Format}
\bibliography{Xu_CHI2026}

\end{document}